\documentclass[11pt,twoside]{article}
\usepackage{asp2014}

\aspSuppressVolSlug
\resetcounters

\begin{document}

\title{Recent advances in the theoretical modeling of pulsating low-mass 
He-core white dwarfs}
\author{A. H. C\'orsico,$^{1}$ L. G. Althaus, $^{1}$ L. M. Calcaferro, 
$^{1}$ A. M. Serenelli,$^2$ S. O. Kepler,$^3$ and C. S. Jeffery$^4$
\affil{$^1$Facultad de Ciencias Astron\'omicas y Geof\'isicas (UNLP), La Plata,  
Argentina; \email{acorsico@fcaglp.unlp.edu.ar}}
\affil{$^2$Institute of Space Sciences (IEEC-CSIC), Barcelona, Spain;}
\affil{$^3$Instituto de F\'isica (UFRGS), Porto-Alegre, Brazil;}
\affil{$^4$Armagh Observatory, College Hill, UK}}

\paperauthor{Sample~Author1}{Author1Email@email.edu}{ORCID_Or_Blank}{Author1 Institution}{Author1 Department}{City}{State/Province}{Postal Code}{Country}
\paperauthor{Sample~Author2}{Author2Email@email.edu}{ORCID_Or_Blank}{Author2 Institution}{Author2 Department}{City}{State/Province}{Postal Code}{Country}
\paperauthor{Sample~Author3}{Author3Email@email.edu}{ORCID_Or_Blank}{Author3 Institution}{Author3 Department}{City}{State/Province}{Postal Code}{Country}

\begin{abstract}
Many extremely low-mass (ELM)  white-dwarf (WD) stars are currently
being found in the field of the Milky Way. Some of these stars exhibit
long-period nonradial $g$-mode pulsations, and constitute the class of ELMV
pulsating WDs. In addition, several  low-mass pre-WDs, which could be
precursors of ELM WDs, have been observed to show short-period 
photometric variations likely due to nonradial $p$ modes and radial modes.
They could constitute a new class of pulsating low-mass pre-WD
stars,  the pre-ELMV stars. Here,  we present the recent results of a
thorough theoretical study of the nonadiabatic pulsation properties of
low-mass He-core WDs and pre-WDs on the basis of fully evolutionary
models representative  of  these stars.
\end{abstract}

\section{Introduction}

An increasing number of low-mass  WDs, including ELM WDs
($M_{\star}\sim 0.18-0.20 M_{\sun}$, H-rich atmospheres),   are
currently being detected through the  ELM survey (see Brown et al.
2016 and references therein).  These WD stars, which likely harbor
cores made of He, are thought to be the result of  strong  mass-loss
events at the red giant branch stage of low-mass stars in binary
systems  before the He flash onset that, in this way, is avoided
(Althaus et al. 2013; Istrate et al. 2016). The increasing interest
in ELM WDs has lead to
the discovery of long-period  ($\Pi \sim 1000-6300$ s)  $g$-mode
pulsations in some of them  (ELMVs).  The existence of ELMV stars ($7000
\lesssim T_{\rm eff} \lesssim 10\,000$ K and $6 \lesssim \log g
\lesssim 7$; red circles in Fig. \ref{Fig1}) constitutes an unprecedented
opportunity   for probing their subsurface layers and ultimately to
place constraints on the currently accepted formation scenarios  by
means  of asteroseismology (Winget \& Kepler 2008; Fontaine \&
Brassard 2008; Althaus et al. 2010). Apart from ELMVs, short-period
($\Pi \sim 300-800$ s)  pulsations in five  objects that are probably
precursors of ELM WDs have been detected  in the few last years.
These stars have
typically $8000 \lesssim T_{\rm eff} \lesssim 13\,000$ K and $4
\lesssim \log g \lesssim 5.5$ (green circles in Fig. \ref{Fig1}) and show  a
surface composition made of H and He. They are called   pre-ELMV stars
and constitute a new class of pulsating stars.  Also, the 
discovery of long-period  ($\Pi
\sim 1600-4700$ s) pulsations in three additional objects located at
the same region of the HR  diagram has been reported. 
These stars are emphasized with black
squares surrounding the green  circles in Fig. \ref{Fig1}. The nature
of these objects is uncertain and  could be identified as pre-ELM
stars as well as SX Phe and/or $\delta$ Scuti  pulsating stars. In
Table \ref{tabla1} we include an updated compilation of the effective
temperature, gravity and range of observed periods for  all the known
pre-ELMV and ELMV stars. 

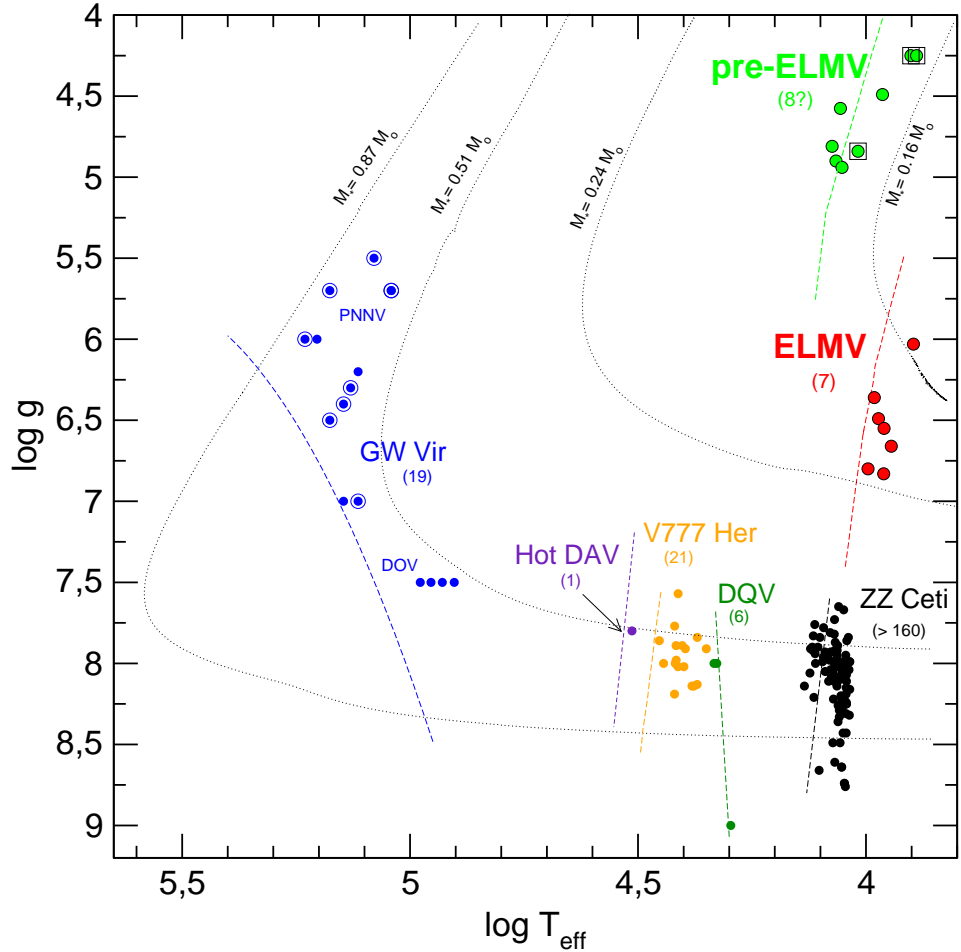
\articlefigure{Corsico_A_Fig1.eps}{Fig1}{The location of the known ELMVs (red
circles) and  pre-ELMVs (light green circles) along with the   other 
several classes of pulsating WD
stars (dots of different colors)   in the   $\log T_{\rm eff} - \log
g$ plane. The three stars emphasized with squares
surrounding the light green  circles can be
identified as pre-ELMV stars as well as SX Phe and/or $\delta$ Scuti
stars.  In parenthesis we include the number of known members
of each class.  Two post-VLTP (Very Late Thermal Pulse) evolutionary
tracks  for H-deficient WDs and two  evolutionary tracks for
low-mass He-core WDs  are  plotted   for reference.  Dashed lines indicate
the theoretical blue edge for the different classes of pulsating WDs.}  

\begin{table}[!ht]
\caption{Stellar parameters and observed period range of the 
known pre-ELMV (upper half of the table)  and ELMV (lower half of the 
table) stars. For ELMVs, the $T_{\rm eff}$ and $\log g$ values are 
computed with 1D model atmospheres after 3D corrections.}
\smallskip
\begin{center}
{\small
\begin{tabular}{lccccr}
\tableline
\noalign{\smallskip}
 Star &  $T_{\rm eff}$ & $\log g$  & $M_{\star}$  & Period range  &  Ref.\\
      &          [K] &    [cgs]  & [$M_{\sun}$] &  [s]          &     \\
\noalign{\smallskip}
\tableline
\noalign{\smallskip}
SDSS J115734.46+054645.6   & $11\,870\pm260$ & $4.81\pm0.13$   & $0.186$ & $364$     & (3) \\  
SDSS J075610.71+670424.7   & $11\,640\pm250$ & $4.90\pm0.14$   & $0.181$ & $521-587$ & (3) \\
WASP J024743.37$-$251549.2 & $11\,380\pm400$ & $4.576\pm0.011$ & $0.186$ & $380-420$ & (1) \\
SDSS J114155.56+385003.0   & $11\,290\pm210$ & $4.94\pm0.10$   & $0.177$ & $325-368$ & (3) \\
KIC 9164561(*)             & $10\,650\pm200$   & $4.86\pm0.04$   & $0.213$ & $3018-4668$ & (5) \\   
WASP J162842.31+101416.7   & $9200\pm600$    & $4.49\pm0.05$   & $0.135$ & $668-755$ & (2) \\
SDSS J173001.94+070600.25(*)  & $7972\pm200$    & $4.25\pm0.5$    & $0.171$ & $3367$      & (4) \\
SDSS J145847.02+070754.46(*)  & $7925\pm200$    & $4.25\pm0.5$    & $0.171$ & $1634-3279$ & (4) \\
\tableline
\noalign{\smallskip}
SDSS J222859.93+362359.6  &  $7890\pm120$  & $5.78\pm0.08$ & $0.142$ & $3254-6235$ &  (6) \\
SDSS J161431.28+191219.4  &  $8700\pm170$  & $6.32\pm0.13$ & $0.172$ & $1184-1263$ &  (6) \\
PSR  J173853.96+033310.8  &  $8910\pm150$  & $6.30\pm0.10$ & $0.172$ & $1788-3057$ &  (8) \\
SDSS J161831.69+385415.15 &  $8965\pm120$  & $6.54\pm0.14$ & $0.179$ & $2543-6125$ &  (9) \\
SDSS J184037.78+642312.3  &  $9120\pm140$  & $6.34\pm0.05$ & $0.177$ & $2094-4890$ &  (10) \\
SDSS J111215.82+111745.0  &  $9240\pm140$  & $6.17\pm0.06$ & $0.169$ &  $108-2855$ &  (7) \\
SDSS J151826.68+065813.2  &  $9650\pm140$  & $6.68\pm0.05$ & $0.197$ & $1335-3848$ &  (7) \\ 
\noalign{\smallskip}
\tableline
\end{tabular}
}
{\footnotesize References: 
(*) Not secure identification as pre-ELM WD (see text);
(1) Maxted et al. (2013); 
(2) Maxted et al. (2014); 
(3) Gianninas et al. (2016); 
(4) Corti et al. (2016); 
(5) Zhang et al. (2016); 
(6) Hermes et al. (2013b); 
(7) Hermes et al. (2013a);
(8) Kilic et al. (2015); 
(9) Bell et al. (2015);
(10) Hermes et al. (2012)
}
\label{tabla1}
\end{center}
\end{table}  

\section{Evolution and pulsation modeling}

We employed stellar evolution models representative of ELM pre-WDs and WDs
generated with the
{\tt LPCODE} stellar evolution code (see Althaus et al. 2013
and references therein).  We carried out
a pulsation stability analysis of  radial and nonradial
$p$ and $g$ modes     employing the nonadiabatic
versions of the  {\tt LP-PUL} pulsation code described in detail  in
C\'orsico et al. (2006). Our nonadiabatic  computations   rely on the
frozen-convection (FC) approximation,  in which  the  perturbation of
the convective flux is neglected. This assumption could constitute  a
source of uncertainties in the location of the derived blue edges of
instability. The equilibrium models employed in our analysis are
realistic  configurations   for low-mass He-core pre-WDs and WDs  
computed by Althaus et al. (2013) by mimicking the binary evolution  
of progenitor stars. Binary evolution was assumed to be  fully
nonconservative, and the loss of angular momentum due to  mass loss,
gravitational wave radiation, and magnetic braking was considered,
following the formalism of Sarna et al. (2000). All
of the He-core pre-WD initial models were derived  from evolutionary
calculations for binary systems consisting of an evolving Main
Sequence low-mass component (donor star)  of initially $1 M_{\sun}$
and a $1.4 M_{\sun}$ neutron star companion  as the other component.
A total of 14 initial He-core pre-WD models  with stellar masses
between  $0.1554$ and $0.4352 M_{\sun}$ were computed for initial
orbital periods at the beginning of the Roche lobe phase in the range
$0.9$ to $300$ d.   

\articlefiguretwo{Corsico_A_Fig2a.eps}{Corsico_A_Fig2b.eps} {Fig2} 
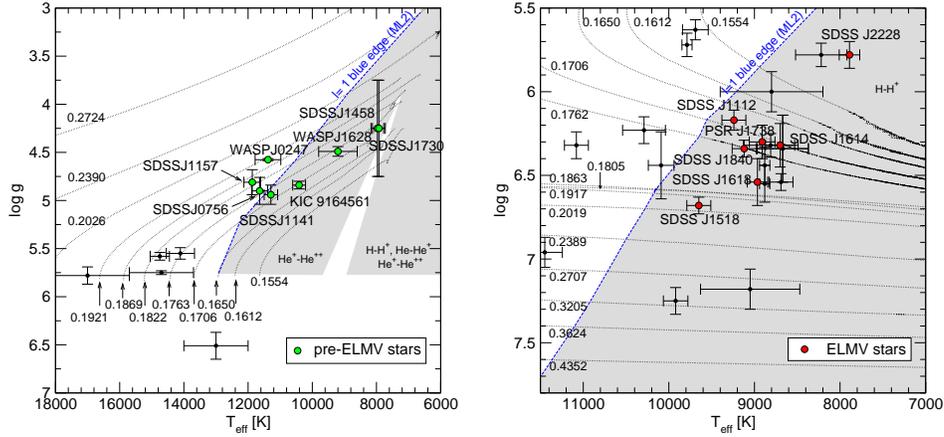
{ \emph{Left:} The
  $T_{\rm eff} - \log g$ diagram showing our  low-mass He-core  pre-WD
  evolutionary tracks (dotted curves) computed neglecting element
  diffusion. Numbers correspond to the stellar mass of each sequence.
  Green dots with error bars correspond to the pre-ELMV stars, and
  black dots depict  the location of stars not observed to vary.
  The dashed
  blue line indicates the nonradial dipole ($\ell= 1$)  blue edge of
  the pre-ELMV instability domain  (emphasized as a gray area) due to
  the $\kappa-\gamma$ mechanism  acting at the He$^+-$He$^{++}$
  partial ionization   region,  as obtained by C\'orsico et
  al. (2016). \emph{Right:} The same diagram but for our low-mass
  He-core WD evolutionary tracks (final cooling branches).  The
  locations of the seven known ELMVs are marked with red circles
  ($T_{\rm eff}$ and $\log g$ computed with 1D model atmospheres after
  3D corrections).  The gray region bounded by the dashed blue line
  corresponds to the instability  domain of $\ell= 1$ $g$ modes due to
  the $\kappa-\gamma$  mechanism acting at the H-H$^{+}$ partial
  ionization region  according to C\'orsico \& Althaus (2016).}

\subsection{Pre-ELMVs}

A complete description of the results presented in this section can be
found in C\'orsico et al.  (2016). Here, we mention only the main
nonadiabatic results.  We analyzed the stability properties
of He-core, low-mass pre-WD models computed assuming the ML2
prescription for the MLT theory of convection (Tassoul et al. 1990) 
and covering a range of
effective temperatures of $25\,000\ {\rm K} \gtrsim T_{\rm eff}
\gtrsim 6000$ K and a range of stellar masses of $0.1554 \lesssim
M_{\star}/M_{\sun} \lesssim 0.2724$. First, we report on results of
evolutionary computations that neglect the action of element diffusion.
For each model, we assessed the pulsational stability of
radial $(\ell= 0)$, and nonradial $(\ell= 1, 2)$ $p$ and $g$ modes
with periods in the range $10\ {\rm s} \lesssim \Pi \lesssim 20\,000$
s.  In the left panel of Fig. \ref{Fig2} we show a spectroscopic HR
diagram ($T_{\rm eff} - \log g$) showing our low-mass He-core pre-WD
evolutionary tracks (dotted curves). Green circles correspond to the
known pre-ELMV stars, and black dots depict the location of
stars not observed to vary. 
The dashed blue line indicates the nonradial $\ell= 1$ $p$-mode
blue edge of the pre-ELMV instability domain (gray area) due
to the $\kappa-\gamma$ mechanism acting at the He$^+$-He$^{++}$
partial ionization region ($\log T \sim 4.7$), as obtained in
C\'orsico et al. (2016).  Our results are in good 
agreement with the predictions of the stability analysis carried 
out by Jeffery \& Saio (2013). 
At $T_{\rm eff} \lesssim 7800$ K there is
also a non-negligible contribution to mode driving due to the
He-He$^{+}$ and H-H$^{+}$ partial ionization zones  ($\log T \sim
4.42$ and $\log T \sim 4.15$, respectively).  The blue edge for
$\ell=2$ modes (not shown) is slightly ($\sim 10-30$ K)
hotter than the $\ell= 1$ blue
edge. The location of the blue edges does not depend on the
prescription for the MLT theory of convection adopted in the
equilibrium models. This is at variance with what happens in the case
of ELMVs (see the next Section). The blue edge of radial modes is
substantially cooler ($\sim 1000$ K) than for nonradial modes. 
Our computations seem to account for the existence
of some of  the known pre-ELMVs, including the observed ranges of  excited
periods. However, our analysis is not able to predict  pulsations in
three pulsating objects (the hottest ones,  see Fig. \ref{Fig2}).
This is directly related to the He abundance ($X_{\rm He}$)  at  the
driving region.  Higher He abundances at the envelope  of our models
would be required in order to the blue edge be hotter, as to include
these three stars within the  instability domain. This could be
achieved by adopting different  masses for the initial donor star in
the original binary system,  that could led to low-mass pre-WD models
with different He abundances at  their envelopes. This issue will be
explored in a future study.

When we take into account the action of  element diffusion
(due to gravitational settling and chemical and thermal diffusion),
$X_{\rm He}$ decreases rapidly at the driving region by virtue of
gravitational settling, and thus the instability region
(not shown) in the $T_{\rm eff} - \log g$ plane becomes
much narrower than in the case of non diffusion, so that none of the
observed pulsating pre-ELMV stars is within that region. This
important result  suggests that \emph{element diffusion could not be
  operative in the pre-WD stage}.  Several factors could be playing a
role to prevent  the effects of element diffusion, namely stellar
winds and/or stellar  rotation. This interesting topic has been
recently addressed by  Istrate \& Fontaine (these proceedings), who
found that rotational  mixing is indeed able to  suppress the effects
of element diffusion. 

\subsection{ELMVs}

We have analyzed the stability properties of about 7000
stellar models of He-core, low-mass WDs with H-pure atmospheres taking
into account  three different prescriptions for the  MLT theory of
convection  (ML1, ML2, ML3; see Tassoul et al. 1990 for their
definitions) and covering a range
of effective  temperatures of   $13\,000\ {\rm K} \lesssim  T_{\rm
  eff} \lesssim 6\,000$ K and a range of stellar masses of $0.1554
\lesssim M_{\star}/M_{\sun} \lesssim 0.4352$.  For each model, we
assessed the pulsation stability of radial ($\ell= 0$) and nonradial
($\ell= 1, 2$) $p$ and $g$ modes with periods  from a range $10\ {\rm
  s} \lesssim \Pi \lesssim 18\,000$ s for the sequence with $M_{\star}= 0.1554
M_{\sun}$, up to a range of periods of $0.3\ {\rm s} \lesssim \Pi \lesssim
5\,000$ s  for the sequence of with $M_{\star}= 0.4352 M_{\sun}$. 
Full details of these calculations are given in C\'orsico \& Althaus (2016).
We show in the right  panel of Fig. \ref{Fig2} the spectroscopic HR
diagram for our low-mass He-core WD evolutionary tracks (final cooling
branches),  along with the location of the seven known ELMVs (red
circles), where  $T_{\rm eff}$ and $\log g$ have been computed with 1D
model atmospheres  after 3D corrections.  The instability domain of
$\ell= 1$ $g$ modes due to the $\kappa-\gamma$  mechanism acting at
the H-H$^{+}$ partial ionization region  is emphasized  with a 
gray region bounded by a dashed
blue line corresponding  to the blue edge of the  instability
domain. Our results are in good agreement with those of  C\'orsico et
al. (2012) and  Van Grootel et al. (2013). Some short-period $g$ modes
are destabilized mainly  by the $\varepsilon$ mechanism due to stable
nuclear burning at the basis of the  H envelope, particularly for
model  sequences with $M_{\star} \lesssim 0.18 M_{\sun}$ (see
C\'orsico \& Althaus 2014a for details).  The blue edge of the
instability domain in the $T_{\rm eff}-\log g$ plane is hotter for
higher stellar mass and larger convective efficiency.  The ML2 and ML3
versions of the MLT theory of convection are the only ones that
correctly account for the location of the seven known ELMV
stars. There is no dependence of the blue edge  of $p$ modes on the
harmonic degree. In the case of $g$ modes, we found a weak sensitivity
of the blue edge with $\ell$. Finally,  the blue edges corresponding
to radial and nonradial $p$ modes are  somewhat ($\sim 200$ K)  hotter
than the blue edges of $g$ modes. We compared the ranges of unstable
mode  periods predicted by our stability  analysis with the ranges of
periods observed in the ELMV stars. We generally found an excellent
agreement.

\section{Conclusions}

The origin and basic nature of pulsations exhibited by pre-ELMVs and
ELMVs  have been established (Steinfadt et al. 2010; 
C\'orsico et al. 2012; Jeffery \& Saio
2013;  Van Grootel et al. 2013; C\'orsico \& Althaus et al. 2014ab,
2016;  C\'orsico et al.  2016; Gianninas et al. 2016). The next step
is to start  exploiting the period spectra of these stars with
asteroseismological analysis.  Asteroseismology
of low-mass He-core WDs  will provide crucial information about the
internal structure and evolutionary status of these stars, allowing us
to place  constraints on the binary evolutionary processes involved in
their formation.

\acknowledgements A.H.C.   warmly thanks  the 
Local Organising Committee  of the  20th European White Dwarf Workshop
for support that allowed him to attend this conference.


\end{document}